\documentclass{article}
\usepackage[T1]{fontenc}
\usepackage[utf8]{inputenc}
\usepackage{ismir} % Remove the "submission" option for camera-ready version
\usepackage{amsmath,cite,url}
\usepackage{graphicx}
\usepackage{color}
\usepackage[inline]{enumitem}
\usepackage{tabularx}
\usepackage{booktabs}
\usepackage{pifont}
\newcommand{\cmark}{\ding{51}}
\newcommand{\xmark}{\ding{55}}
\usepackage{multirow}

\NewDocumentEnvironment{inlinelist}{O{label=(\roman*)}}
  {\begin{enumerate*}[#1]}
  {\end{enumerate*}}

\newcommand{\modelsNumber}{26}
\newcommand{\libraryName}{MxlTok}
\newcommand{\participantsNumber}{16}

\newcommand{\ap}[1]{\textcolor{black}{#1}}

\title{Does Notation Matter? MusicXML Tokenisation for Symbolic Music Generation and Understanding}

\multauthor
  {Andrea Poltronieri$^1$ \hspace{1cm} Matteo Spanio$^2$ \hspace{1cm} Simone Chieppa$^1$}
  {{\bf Xavier Serra$^1$ \hspace{1cm} Martín Rocamora$^1$}\\
  $^1$ Music Technology Group, Universitat Pompeu Fabra, Barcelona, Spain\\
  $^2$ University of Padova, Padova, Italy\\
  {\tt\small andrea.poltronieri@upf.edu, spanio@dei.unipd.it}
}

\def\authorname{A. Poltronieri, M. Spanio, S. Chieppa, X. Serra, and M. Rocamora}

\usepackage[bookmarks=false,pdfauthor={\authorname},pdfsubject={\pdfsubject},hidelinks]{hyperref}
\begin{document}

\maketitle

\begin{abstract}
Symbolic music representations encode musical content at varying levels of completeness, from performance-oriented formats such as MIDI to fully notated scores such as MusicXML. 
Yet the impact of this representational richness on deep learning models remains poorly understood.
While tokenisation has been extensively studied for MIDI, score-based formats have received considerably less attention, and MusicXML in particular remains largely unexplored despite being the de-facto standard for score interchange.
In this paper, we introduce the first general-purpose tokenisation framework for MusicXML, and conduct a systematic empirical study to determine whether notational richness confers a measurable advantage for deep learning on symbolic music.
We train \modelsNumber\ autoregressive and masked language models, comparing MusicXML tokenisation against MIDI and ABC-based representations, probing the effect of different tokenisation strategies within MusicXML, and testing whether music-grammar-aware tokenisation outperforms treating scores as plain text.
We evaluate music understanding tasks using linear probing across composer classification, difficulty estimation, and accompaniment suggestion, and generative tasks using objective statistical metrics and a human pairwise preference study.
Our results show that MusicXML-tokenised models outperform MIDI and text-based counterparts on understanding and generative tasks.
\end{abstract}

\section{Introduction}
\label{sec:intro}

% - representing music is challenging
%     - audio and symbolic representations
%     - symbolic representation systems
%     - expressiveness vs. completeness
%     - performance vs. score representations (descriptive vs. prescriptive)

\begin{figure}[t]
\centering
% Score image (centered)
\includegraphics[width=0.38\columnwidth]{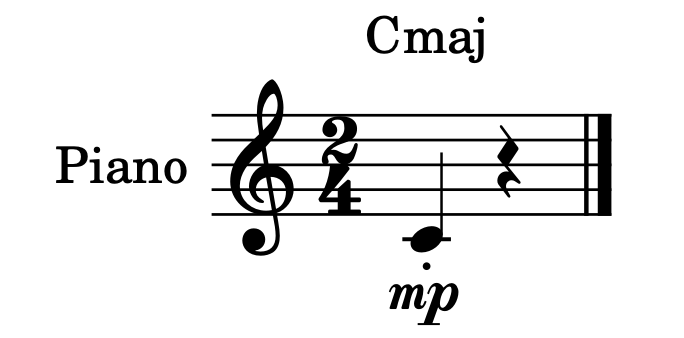}

\vspace{-0.5em}

% Tokenized sequences (left-aligned, same height)
\begin{flushleft}
\includegraphics[height=1.61cm]{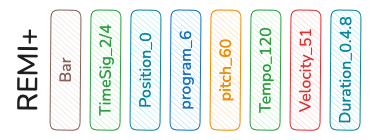}

\vspace{0.3em}

\includegraphics[height=1.61cm]{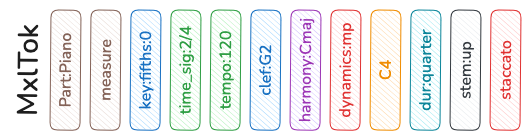}
\end{flushleft}

\vspace{-1.5em}
\caption{Tokenisation of a C4 staccato quarter note over a C major chord under REMI+ and \libraryName. 
Unlike REMI+, \libraryName\ encodes harmonic and notational context.}
\label{fig:tokenization}
\vspace{-1em}
\end{figure}

Music can be computationally represented at multiple levels~\cite{vinet2004levels}, from raw audio signals to symbolic encodings of discrete musical events--such as notes, chords, and rhythms--that support structural analysis, manipulation, and retrieval~\cite{marsden2005representation}.
Within the symbolic domain, however, a fundamental tension persists between \emph{expressive completeness} and \emph{structural generality}~\cite{wiggins1993framework}: the capacity to faithfully reconstruct the original content on one hand, and the ability to represent and manipulate high-level musical structures on the other. 
This trade-off is further complicated by the distinction between \emph{prescriptive} and \emph{descriptive} representations~\cite{seeger1958music-writing}: score-based formats specify what should be played, abstracting the expressive nuances of a performance, while performance-oriented formats capture what was played, at the cost of notational generality.

These two axes---completeness versus generality, and prescriptive versus descriptive---largely define the landscape of symbolic formats in use today.
Performance-oriented representations, such as MIDI~\cite{rothstein1992midi}, encode how music was played---pitch, velocity, and timing.
% ---but exhibit several limitations from a notational standpoint: time is represented as absolute ticks rather than musical units such as beats and measures; enharmonic equivalents are collapsed into a single pitch class, losing spelling information; voices and parts are flattened into a single event stream with no hierarchical score 
% structure; and expressive markings such as dynamics and articulations are either absent or reduced to coarse velocity approximations~\cite{wiggins2000musical-knowledge}.
\ap{Some notational categories can be recovered from this data: metrical position through quantisation, dynamic and articulation markings from velocity and duration. Others are lost entirely: enharmonic equivalents collapse into a single pitch class, and score structure is flattened into events and tracks with no hierarchy of parts, staves, and voices~\cite{wiggins2000musical-knowledge}.}
% Score-based formats instead encode these elements explicitly, and are prescriptive by design.
Score-based formats, by contrast, are prescriptive by design, explicitly encoding these elements.
ABC notation~\cite{walshaw2011abc} has seen growing adoption for score-based deep learning~\cite{qu2024mupt,wang2025notagen}, where scores are typically treated as plain text.
% MusicXML, the de-facto standard for score interchange, offers a richer representation---capturing enharmonic spelling, multi-voice polyphony, articulations, dynamics, and hierarchically organised parts and measures---yet has remained largely overlooked in the MIR deep learning literature~\cite{mayer2024lmx,suzuki2023mxl-tokenizer}.
MusicXML, the de-facto standard for score interchange, offers a richer representation---capturing enharmonic spelling, multi-voice polyphony, articulations, dynamics, and hierarchically organised parts and measures---yet remains largely unexplored in the MIR literature~\cite{mayer2024lmx}.

Applying symbolic representations in deep learning requires converting them into discrete token sequences, a process that has been extensively studied for MIDI~\cite{fradet2021miditok,fradet2023bpe,fradet2023impact}, yet has received considerably less attention for score-based formats.
We address this gap by introducing \libraryName, the first general-purpose tokenisation framework for MusicXML, and conducting a systematic empirical study of whether notational richness confers a measurable advantage for deep learning on symbolic music.
To this end, we train \modelsNumber\ models spanning two architectures---autoregressive and masked language models---and evaluate them on both generative and understanding tasks.
We systematically compare MusicXML tokenisation against MIDI and ABC representations, probing the effect of serialisation strategy and vocabulary size within MusicXML, and testing whether music-informed tokenisation outperforms plain-text encoding~\cite{qu2024mupt,wang2025notagen}.
% To this end, we train \modelsNumber\ models spanning two architectures--autoregressive and masked language models--and evaluate them on both generative and understanding tasks, comparing MusicXML tokenisation against MIDI and ABC-based representations, probing the effect of serialisation strategy within MusicXML, and testing whether music-grammar-aware tokenisation outperforms treating scores as plain text, as is common in recent work~\cite{qu2024mupt,wang2025notagen}.
Our results suggest that the expressiveness of MusicXML is particularly advantageous for understanding tasks, where MusicXML-tokenised models consistently outperform their MIDI and text-based counterparts in all probing benchmarks.
Moreover, MusicXML tokenisation performs competitively on generative tasks, achieving better scores in objective statistical metrics and ranking highest in human preference evaluations, while producing fully notated scores that encode expressive nuances like dynamics and articulations.

The contributions of this paper can be summarised as follows:
\begin{inlinelist}
\item \textbf{\libraryName}: We introduce the first general-purpose tokenisation framework for MusicXML,
% , distributed as an open-source Python library.
following the extensible design of MidiTok~\cite{fradet2021miditok}, with built-in support for custom tokenisation strategies, Byte Pair Encoding (BPE) training~\cite{sennrich2016bpe,fradet2023bpe}, and dataset preprocessing.
\item \textbf{Empirical study}: We train \modelsNumber\ models to systematically compare tokenisation strategies along three axes: serialisation order within MusicXML (measure-first vs. part-first); symbolic input representation (MusicXML vs. MIDI vs. ABC notation); and tokens grammar (music-specific vs. general-purpose subword tokenisation).
All comparisons are conducted across both autoregressive and masked language model architectures, with and without BPE.
% , holding model architectures fixed to isolate the effect of the input representation.
\item \textbf{Evaluation protocol}: Understanding tasks are assessed via linear probing on contextual embeddings extracted from trained masked language models, across composer classification, difficulty estimation, and accompaniment suggestion.
Generative tasks are evaluated using objective statistical metrics~\cite{dong2020muspy}, alongside a human pairwise preference study~\cite{chiang2024chatbot, kim2025music}.
\item \textbf{Open models}: We release the weights of 7 pretrained MusicXML-based models---both generative and understanding---covering all \libraryName\ tokenisation configurations. 
To our knowledge, these are the first general-purpose and publicly available language models trained on MusicXML data.\footnote{\url{https://github.com/andreamust/MxlTok}.}
\end{inlinelist}

\section{Related Work}
\label{sec:sota}

\subsection{MIDI Tokenisers}
\label{ssec:sota:symbolic-tokenizers}

Tokenisation of MIDI has been studied extensively through a variety of strategies, unified in the MidiTok library~\cite{fradet2021miditok}, ranging from direct MIDI event encoding~\cite{oore2018thistime} to musically structured representations with explicit bar and duration tokens~\cite{huang2020pop,vonrutte2023figaro}, and embedding pooling approaches to reduce sequence length~\cite{hsiao2021compound,zeng2021musicbert,ren2020popmag,ens2020mmm}.
Among these, REMI+~\cite{vonrutte2023figaro} and Octuple~\cite{zeng2021musicbert} have emerged as the most widely adopted~\cite{fradet2023bpe, fradet2023impact}, and we use them as baselines in our empirical study.
REMI+ extends REMI~\cite{huang2020pop} to the multitrack, multi-time-signature setting by prepending \texttt{Program} tokens to each note event, while Octuple reduces sequence length by merging multiple note attributes into a single pooled embedding, at the cost of requiring multiple output heads during training \ap{in its original formulation}.
More recently, Qian et al.~\cite{qian2026pianoroll} propose Pianoroll-Event, which encodes scores as structured pianoroll event sequences, prioritising encoding efficiency over notational completeness.

Beyond the design of tokenisation schemes, two lines of work are directly relevant to our study.
Fradet et al.~\cite{fradet2023bpe} apply byte pair encoding to symbolic music tokenisation, showing that learned subword-like merges improve performance on both generative and understanding tasks---a setup closely related to ours, and one that motivates our inclusion of BPE as a variable across all experimental conditions.
\ap{BPE~\cite{sennrich2016bpe} is a subword tokenisation algorithm that iteratively merges the most frequent token pair into a new vocabulary entry until a target size is reached.}
% : in \texttt{aaabdaaabac}, \texttt{aa} is first merged into a single token, yielding \texttt{ZabdZabac}.}
A subsequent study~\cite{fradet2023impact} analyses the effect of time and note duration representations on model performance across composer classification, emotion classification, music generation, and sequence representation tasks, finding that explicit encoding of temporal information yields task-dependent benefits; however, the study is restricted to MIDI and offers limited generalisable conclusions across tokenisation families.

\ap{Both studies, however, are restricted to MIDI, leaving open whether symbolic format choice or tokenisation granularity is the primary driver of model performance. 
Comparisons across formats, using hand-crafted features~\cite{simonetta2023optimizing} or neural representations~\cite{zhang2023classification}, report no consistent advantage for notationally richer ones; a survey of both~\cite{simonetta2025style} attributes this to tooling rather than to the information itself. In this study, we make tokenisation the controlled variable.}
% Both studies, however, are restricted to MIDI, leaving open whether symbolic format choice or tokenisation granularity is the primary driver of model performance, a question this paper directly addresses.
% To our knowledge, no equivalent study exists for score-based formats, leaving open the question of whether symbolic format choice or tokenisation granularity is the primary driver of model performance.
% To our knowledge, no equivalent study exists for score-based formats, leaving open the question of whether the choice of symbolic format or the granularity of the music features being tokenised is the primary driver of model performance.

\subsection{Score Tokenisers}
\label{ssec:sota:score-tokenizers}

Compared to MIDI, tokenisation of score-based formats has received considerably less attention.
The main approaches operating on score representations are generative models based on ABC notation~\cite{qu2024mupt,wang2025notagen}.
However, ABC was originally designed for encoding folk and traditional Western European tunes, which are 
typically single-voice melodies written on a single staff.
The standard has not been officially updated since version 2.1~\cite{walshaw2011abc}, and brings some limitations: multi-voice support remains inconsistently implemented across tools, and both MuPT~\cite{qu2024mupt} and NotaGen~\cite{wang2025notagen} introduce non-standard extensions to handle polyphonic content, resulting in mutually incompatible dialects that hinder reproducibility and interoperability.

From a tokenisation perspective, both works treat ABC scores as plain character sequences, applying general-purpose subword tokenisers without exploiting any music-specific grammatical structure.
As a result, musical concepts are not represented as atomic units: for example, a C7 note is encoded as the character sequence \texttt{c''}, where the pitch is implied by the combination of the letters, rather than being represented as a single explicit token.

\begin{table}[t]
\centering
\footnotesize
\begin{tabularx}{\columnwidth}{X c c c c}
\toprule
\textbf{Feature} & \textbf{MIDI} & \textbf{ABC} & \textbf{LMX} & \textbf{\libraryName} \\
\midrule
Enharmonic spelling  & \xmark & \cmark & \cmark & \cmark \\
Dynamics \& hairpins           & $\sim$ & \cmark & \xmark & \cmark \\
Tempo markings                 & $\sim$ & \cmark & \xmark & \cmark \\
Articulation \& ornaments      & \xmark & $\sim$ & \xmark & \cmark \\
Harmony / chord symbols        & $\sim$ & $\sim$ & \xmark & \cmark \\
Multi-part score structure     & $\sim$ & $\sim$ & \xmark & \cmark \\
Repeat signs                   & \xmark & \cmark & \xmark & \cmark \\
\bottomrule
\end{tabularx}
\caption{Notational features supported by each symbolic representation. 
\cmark: fully supported; $\sim$: partially supported; \xmark: not supported.}
\label{tab:features}
\vspace{-1em}

\end{table}

Tokenising score-based formats such as MusicXML naturally provides an explicit and hierarchical musical grammar, since the format itself organises musical content into discrete, semantically meaningful elements---pitch, duration, articulation, dynamics, voice, harmony---structured across parts, measures, and notes, rather than encoding them as a flat character stream.
Only a handful of approaches have attempted to exploit this structure, and all are task-specific and notionally incomplete.
Suzuki~\cite{suzuki2023mxl-tokenizer} introduces a MusicXML tokeniser for piano score rearrangement, covering basic notational elements but restricted to piano and lacking support for expressive markings.
Mayer et al.~\cite{mayer2024lmx} propose LMX, a depth-first serialisation of the MusicXML tree for optical music recognition, which explicitly discards dynamics, pedal markings, and does not support multi-part scores or repetitions.
\ap{Zhang et al. \cite{zhang2023classification} adapt REMI and Compound Word to MusicXML by introducing key-signature and voice tokens, but retain the underlying MIDI event grammar and evaluate only on piece-level classification.}

\libraryName\ addresses these limitations by providing the first general-purpose MusicXML tokenisation framework, encoding the full range of notational elements available in the source format (Table~\ref{tab:features}).

\section{Methodology}
\label{sec:methodology}

\subsection{The \libraryName\ Framework}
\label{ssec:method:tokenizer}

\libraryName\ is designed as a general-purpose tokenisation framework for MusicXML, extending the linearisation approach of LMX~\cite{mayer2024lmx} to support the full set of notational elements available in the format---including pitch spelling, dynamics and hairpins, tempo markings, articulations and ornaments, harmony symbols, repeat signs, and multi-part scores (Table~\ref{tab:features}).
Following the extensible design of MidiTok~\cite{fradet2021miditok}, \libraryName\ exposes a base tokeniser class from which users can derive custom strategies by overriding format-specific methods, while inheriting shared infrastructure for vocabulary construction, BPE training, and dataset preprocessing.
The library ships with three built-in tokenisers: the default \texttt{mxl} tokeniser, 
and \texttt{REMI+} and \texttt{Octuple}, which apply their original MIDI tokenisation 
schemes to MusicXML-parsed score content.

Each note is represented as an ordered group of tokens consisting of a set of prefix tokens encoding contextual modifiers (e.g., \texttt{grace}, \texttt{chord}), a root token encoding note type and duration (e.g., \texttt{quarter}, \texttt{eighth}), and a set of suffix tokens encoding notational modifiers (e.g., \texttt{dot}, \texttt{tied:start}, dynamics, articulations), as shown in Figure~\ref{fig:tokenization}.
The serialiser applies delta-encoding for voice, stem, and staff tokens, emitting them only when their value changes with respect to the previous event.
% The serialiser applies delta-encoding for voice, stem, and staff tokens: rather than repeating these tokens on every note, they are emitted only when their value changes with respect to the previous event.
% , reducing sequence length by 15--30\% without information loss.
\libraryName\ supports microtonal pitch alterations, enabling application to non-Western music traditions.
\ap{The base vocabulary comprises approximately 750 tokens, against the 200–400 typical of MIDI tokenisers.}

A key design dimension of \libraryName\ is the \emph{serialisation order} of the score, which directly mirrors the two document types defined in the MusicXML standard~\cite{musicxml40}.
In \emph{Part-First} (PF) serialisation, corresponding to \texttt{<score-partwise>}, the score is traversed part by part, with all measures of a given instrument appearing before moving to the next.
In \emph{Measure-First} (MF) serialisation, corresponding to \texttt{<score-timewise>}, measures are primary, and all parts within a given measure are serialised together before advancing to the next.
% , preserving temporal alignment across instruments at the cost of interleaving voices.
\ap{Within a measure, parts are emitted as contiguous blocks; this keeps simultaneous material within a bar's distance in the token stream, at the cost of interleaving voices across parts.}
Both strategies are lossless and encode the same musical content, but impose different inductive biases on the model; their relative merits for generative and understanding tasks are an empirical question addressed in this paper.
BPE is supported at arbitrary vocabulary sizes; in this work we train models with 5k and 10k tokens, independently for each tokeniser configuration, consistent with the approach of Fradet et al.~\cite{fradet2023bpe}.
Beyond these experimental axes, \libraryName\ exposes additional configurable dimensions---including interval-based pitch encoding and pitch-shift augmentation---following the design space explored by previous studies~\cite{fradet2023impact}.

\subsection{Dataset}
\label{ssec:method:dataset}
We train all models on the Rated and Deduplicated (R$\cap$D) subset of PDMX \ap{(\textit{v5})}~\cite{long2024pdmx}, a large-scale dataset of public domain MusicXML scores collected from the MuseScore platform.
All scores in PDMX are released under CC-0 or Public Domain licenses, allowing us to release both the preprocessed tokenised data and the weights of all trained models---including the generative ones---without licensing restrictions, ensuring full reproducibility of our results.

PDMX comprises 254K scores in total, organised into four subsets obtained by progressively filtering for deduplication and user ratings: All (A, 254K scores), Deduplicated (D, 102K), Rated (R, 14K), and Rated and Deduplicated (R$\cap$D, 13K).
Deduplication is performed via embedding-based title similarity and note count clustering, retaining the highest-rated arrangement per cluster; ratings, provided by the MuseScore community, serve as a proxy for data quality.
% We use the R$\cap$D subset as Long et al.~\cite{long2024pdmx} show that, despite being the smallest subset, it consistently achieves the highest scores in both objective metrics and human preference evaluations, outperforming models trained on the full dataset.
We use the R$\cap$D subset following the ablation study of \cite{long2024pdmx}, which shows that rating-based filtering is a more effective quality proxy than scale, with R$\cap$D-trained models outperforming those trained on larger subsets in both objective metrics and human preference evaluations.
The R$\cap$D subset covers a diverse range of genres, with classical and folk forming the largest portion, followed by rock, pop, jazz, and soundtrack.
\ap{Notational content is pervasive rather than incidental in R$\cap$D: 53\% of scores carry dynamics, 50\% articulations, and 61\% at least one note whose spelling is unrecoverable from pitch class.}

% The R$\cap$D subset is dominated by classical and folk music, with a longer tail of genres, including rock, pop, jazz, and soundtrack.
% compared to the unfiltered dataset, reflecting the higher engagement of the MuseScore community with more diverse repertoire.

\subsection{Baselines and Experimental Conditions}
\label{ssec:method:baselines}

\vspace{-0.8em}

\begin{table}[ht]
\centering
\footnotesize
\begin{tabularx}{\columnwidth}{l l l c c}
\toprule
\textbf{Tokeniser} & \textbf{Format} & \textbf{BPE} & \textbf{GPT} & \textbf{BERT} \\
\midrule
\libraryName--PF & MusicXML & none, 5k, 10k & \cmark & \cmark \\
\libraryName--MF & MusicXML & none, 5k, 10k & \cmark & \cmark \\
\libraryName-Octuple   & MusicXML & none          & \xmark & \cmark \\
REMI+                  & MIDI     & none, 5k, 10k & \cmark & \cmark \\
Octuple                & MIDI     & none          & \xmark & \cmark \\
ABC                    & ABC      & none, 5k, 10k & \cmark & \cmark \\
\bottomrule
\end{tabularx}
\caption{Experimental conditions. For each tokeniser, \cmark\ indicates which 
model architectures are trained.}
\label{tab:conditions}
\vspace{-0.5em}

\end{table}

To compare MusicXML tokenisation against alternative symbolic representations, we prepare the R$\cap$D subset of PDMX in each target format prior to tokenisation.
% MIDI files are provided natively in the PDMX dataset~\cite{long2024pdmx}, while ABC conversion is performed using the xml2abc library~\cite{xml2abc}, following prior work~\cite{yang2026legato}.
MIDI files are obtained directly from the PDMX dataset~\cite{long2024pdmx}, while ABC conversion is performed using the xml2abc library~\cite{xml2abc}, following prior work~\cite{yang2026legato}.
Both conversions are lossless with respect to target format, as MusicXML is the richest of the three representations.

For MusicXML, we evaluate two \libraryName\ tokenisers: the default \texttt{mxl} tokeniser under both part-first (PF) and measure-first (MF) serialisation orders, and \libraryName-Octuple.
For MIDI, we use REMI+~\cite{vonrutte2023figaro} and Octuple~\cite{zeng2021musicbert}, the two most widely adopted MIDI tokenisation schemes, shown to achieve the best performance on generative and understanding tasks respectively~\cite{fradet2023bpe,fradet2023impact}.
The ABC baseline tokenises scores as plain text using the HuggingFace BPE tokeniser~\cite{wolf2021huggingface}, consistently with prior work~\cite{qu2024mupt,wang2025notagen}.
BPE is evaluated at vocabulary sizes of 5k and 10k tokens for all tokenisers except Octuple, which is excluded as its embedding pooling design already serves as a sequence compression strategy.
For autoregressive models, every tokenised sequence is prefixed, immediately after the \texttt{<BOS>} token, with the number of parts and their instrument names; this ensures that all models receive the same structural conditioning at generation time, enabling a fair comparison across tokeniser configurations (c.f. Section~\ref{ssec:eval:generation}).
Both Octuple variants are likewise excluded from generative evaluation, as it has been shown to be underperforming on autoregressive generation~\cite{fradet2023bpe}.
All experimental conditions are summarised in Table~\ref{tab:conditions}.

% To compare MusicXML tokenisation against alternative symbolic representations, we implement two baselines using the same model architecture and training procedure described below.
% The MIDI baseline is implemented using MidiTok~\cite{fradet2021miditok}, converting scores to MIDI and tokenising with both REMI+~\cite{vonrutte2023figaro} and Octuple~\cite{zeng2021musicbert}.
% The ABC baseline converts scores to ABC notation and tokenises using the HuggingFace built-in BPE tokeniser~\cite{wolf2021huggingface}, treating the score as plain text consistently with prior work~\cite{qu2024mupt,wang2025notagen}.

\subsection{Models}
\label{ssec:method:models}

All \modelsNumber\ models are trained from scratch using the HuggingFace Transformers library~\cite{wolf2021huggingface}.
% , with architecture held fixed across all conditions to isolate the effect of the input representation.
We train two model families: an autoregressive transformer (GPT-2~\cite{radford2019language}) for generative tasks, and a masked language model (BERT~\cite{devlin2019bert}) for understanding tasks.
Both follow the configuration of Long et al.~\cite{long2024pdmx}: 6 layers, 8 attention heads, and a hidden dimension of 512, totalling approximately 20M parameters.
This architecture was validated by \cite{long2024pdmx} on the same dataset and task setup, achieving strong performance without requiring larger model capacities; we adopt it to ensure that observed differences are attributable to tokenisation rather than architectural choices.
Since BPE and no-BPE models operate on sequences of substantially different lengths, convergence rates vary across configurations; we therefore train on a maximum of 100k steps and apply early stopping with a patience of 1500 steps.
% All models are trained on 4$\times$A100 GPUs (40GB each).
\section{Evaluation}
\label{sec:evaluation}

\begin{table*}[ht]
\centering
\renewcommand{\arraystretch}{1.2}
\resizebox{.9\textwidth}{!}{%
\begin{tabular}{llc cc cc ccc | cc}
\toprule
& & & \multicolumn{2}{c}{\textbf{Composer Classification}} & \multicolumn{2}{c}{\textbf{Difficulty Estimation}} & \multicolumn{3}{c}{\textbf{Accompaniment Suggestion}} & \multicolumn{2}{c}{\textbf{Ablation Study}} \\
\cmidrule(lr){4-5} \cmidrule(lr){6-7} \cmidrule(lr){8-10} \cmidrule(lr){11-12}
\textbf{Format} & \textbf{Tokenizer} & \textbf{BPE}
  & \textbf{Acc} $\uparrow$ & \textbf{Macro F1} $\uparrow$
  & \textbf{Acc-9} $\uparrow$ & \textbf{Acc-$\pm$1} $\uparrow$
  & \textbf{MAP} $\uparrow$ & \textbf{H@1} $\uparrow$ & \textbf{H@5} $\uparrow$
  & \textbf{Acc} $\uparrow$ & \textbf{Macro F1} $\uparrow$ \\
\midrule
\multirow{6}{*}{MusicXML}
  & \multirow{3}{*}{\libraryName\--MF} & 10k  & \textbf{0.781} & \textbf{0.777} & \textbf{0.303} & \textbf{0.695} & 0.779 & 0.650 & 0.949 & 0.735 & 0.732 \\
  &                                    & 5k   & 0.775 & 0.772 & 0.301 & 0.650 & 0.810 & 0.695 & 0.965 & 0.733 & 0.725 \\
  &                                    & None & 0.769 & 0.765 & 0.278 & 0.619 & \textbf{0.815} & \textbf{0.702} & \textbf{0.973} & 0.701 & 0.690 \\
  & \multirow{3}{*}{\libraryName\--PF} & 10k  & 0.753 & 0.747 & 0.277 & 0.659 & 0.777 & 0.652 & 0.952 & 0.675 & 0.668 \\
  &                                    & 5k   & 0.746 & 0.741 & 0.246 & 0.620 & 0.804 & 0.687 & 0.958 & 0.678 & 0.667 \\
  &                                    & None & 0.763 & 0.759 & 0.258 & 0.608 & 0.803 & 0.680 & 0.966 & 0.672 & 0.664 \\
\midrule
\multirow{1}{*}{MusicXML}
  & \multirow{1}{*}{Octuple}           & None & 0.742 & 0.735 & 0.242 & 0.603 & 0.750 & 0.618 & 0.931 & 0.680 & 0.664 \\
\midrule
\multirow{3}{*}{MIDI}
  & \multirow{3}{*}{REMI+}             & 10k  & 0.748 & 0.739 & 0.229 & 0.608 & 0.721 & 0.573 & 0.922 & 0.734 & 0.724 \\
  &                                    & 5k   & 0.744 & 0.735 & 0.228 & 0.551 & 0.726 & 0.579 & 0.925 & 0.743 & 0.729 \\
  &                                    & None & 0.721 & 0.708 & 0.261 & 0.667 & 0.724 & 0.579 & 0.924 & 0.735 & 0.725 \\
\midrule
\multirow{1}{*}{MIDI}
  & \multirow{1}{*}{Octuple}           & None & 0.716 & 0.710 & 0.241 & 0.622 & 0.614 & 0.458 & 0.820 & 0.725 & 0.717 \\
\midrule
\multirow{3}{*}{ABC}
  & \multirow{3}{*}{Word-level}        & 10k  & 0.705 & 0.698 & 0.272 & 0.632 & 0.602 & 0.445 & 0.814 & 0.676 & 0.668 \\
  &                                    & 5k   & 0.715 & 0.707 & 0.256 & 0.633 & 0.595 & 0.432 & 0.811 & 0.674 & 0.666 \\
  &                                    & None & 0.707 & 0.703 & 0.267 & 0.660 & 0.556 & 0.382 & 0.770 & 0.668 & 0.660 \\
\bottomrule
\end{tabular}%
}
\caption{Downstream task evaluation across tokenisation strategies. Best results per column are in bold. Standard deviations are omitted, as all values are below a threshold of 0.05.
}
\label{tab:downstream_results}
\vspace{-1em}
\end{table*}

Following prior work~\cite{fradet2023bpe,fradet2023impact,qian2026pianoroll,zhao2025abc,hachana2025using, alonso2025omar}, we evaluate understanding and generative tasks separately.
Understanding is assessed via linear probing, while generative quality is measured using both objective metrics and a human preference study.

\subsection{Music Understanding Tasks}

% Since no prior work has addressed deep learning on MusicXML at the scale we consider, no established benchmark or dataset exists for evaluating such models.
Since no prior work has addressed deep learning on MusicXML, no established benchmark or dataset exists for evaluating such models.
We propose a set of probing tasks and the corresponding datasets.
For all tasks, embeddings are extracted from each BERT model and used as fixed features for a probing classifier consisting of two linear layers with hidden dimension 512.
% For all tasks, embeddings are extracted by averaging across all layers of each BERT-style model and used as fixed features for a probing classifier consisting of two linear layers with hidden dimension 512.
% , trained for up to 100 epochs with early stopping.
% , designed to assess whether different tokenisation strategies give rise to representations that encode musically meaningful information.

\subsubsection{Composer Recognition}

We evaluate composer classification on a subset of the KernScores corpus~\cite{sapp2005kernscores}, a collection of symbolic music scores in the Humdrum \texttt{**kern} format.
Scores are converted to MusicXML using the MuseScore CLI~\cite{musescore}, a lossless conversion with respect to the musical content.
The dataset comprises 1,968 scores spanning 20 composers with a balanced class distribution; we define balanced splits and evaluate using 5-fold cross-validation.
To isolate the effect of representational richness from tokenisation strategy, we additionally run an ablation in which the same scores are first converted to MIDI and then back to MusicXML and ABC notation, ensuring that all three representations are derived from the same lossy source.

% Embeddings are extracted from the penultimate layer of each BERT-style model and used as fixed features for a probing classifier consisting of two linear layers with hidden dimension 512, trained for up to 100 epochs with early stopping.

\subsubsection{Difficulty Estimation}

For difficulty estimation, we evaluate on the \textit{Can I Play It?} (CIPI) dataset~\cite{ramoneda2024cipi}, which comprises 652 classical piano compositions across 9 difficulty levels and 29 composers from the Baroque era to the 20th century. 
The dataset is distributed in MusicXML format; for MIDI and ABC models, scores are converted as described in Section~\ref{sec:methodology}. 
We adopt the predefined 5-fold cross-validation splits and report 9-class accuracy 
and balanced $\pm$1 accuracy.

\subsubsection{Accompaniment Suggestion}

We adopt the accompaniment suggestion task from \cite{hachana2025using}, evaluated on the Wikifonia dataset~\cite{simonetta2018wikifonia}, which consists of 5,000 lead sheets in MusicXML format.
Positive pairs are constructed by separating the melody track from the remaining accompaniment, while negative pairs use non-matching combinations.
The task is evaluated as a retrieval problem using Mean Average Precision (MAP) and HITS@$k$ ($k = 1, 5$).

\subsection{Generative Tasks}
\label{ssec:eval:generation}

\subsubsection{Objective Analysis}

Evaluating generated music objectively remains an open challenge~\cite{lerch2025survey}. 
While MusPy~\cite{dong2020muspy} provides metrics designed for symbolic music generation across all formats considered here, their correlation with perceived quality remains poorly understood.
% The only partial evidence comes from~\cite{long2024pdmx}, who find a positive correlation between pitch class entropy and score quality, and negative correlations for scale consistency and groove consistency; however, this analysis is confined to highly-rated MuseScore uploads (ratings 4--5), where quality is defined by community ratings rather than perceptual judgements.

Recently,~\cite{qian2026pianoroll} proposed aggregating Polyphony Rate (PR), Scale Consistency (SC), and Groove Consistency (GC) into a single Jensen--Shannon (JS) score, capturing harmonic richness, tonal coherence, and rhythmic stability respectively. 
The score is defined as $\text{JS} = 100 \times \exp\!\left(-2 \times \overline{\text{JS}}\right)$, where $\overline{\text{JS}}$ is the average Jensen--Shannon divergence between the distributions of the generated and reference
sequences for each metric.
Higher values indicate closer alignment between generated and reference distributions.
We adopt this formulation, using statistics computed over the training split of the PDMX R$\cap$D subset as reference distribution, since it proved to be a clean and diverse dataset of MusicXML data~\cite{long2024pdmx}. 

To ensure comparability, all models are conditioned on the same set of musical attributes---number of parts, instrument types, key, and time signature.
% , keeping these factors constant so that any observed differences in the objective metrics reflect tokenization choices rather than differences in the musical material being generated. 
For each model, 200 pieces are generated over 20 prompts with 10 outputs each, using a maximum sequence length of 2048 tokens.
% using 20 prompts with 10 generation each.
% Each prompt specifies the number of instruments, instrument identities, key signature, and time signature, ensuring that all models generate music under identical conditions and that results are directly comparable across tokeniser configurations.

\subsubsection{Human Listening Test}

\begin{table*}[ht!]
\centering
\renewcommand{\arraystretch}{1.2}
\resizebox{.8\textwidth}{!}{%
\begin{tabular}{llc cccc cccc}
\toprule
& & & \multicolumn{4}{c}{\textbf{Statistical Evaluation}} & \multicolumn{4}{c}{\textbf{User Study}} \\
\cmidrule(lr){4-7} \cmidrule(lr){8-11}
\textbf{Format} & \textbf{Tokenizer} & \textbf{BPE}
  & \textbf{PR} & \textbf{SC} & \textbf{GC} & \textbf{JS} $\uparrow$ 
  & \textbf{Fidelity} $\uparrow$  & \textbf{Correctness} $\uparrow$  & \textbf{Diversity} $\uparrow$ & \textbf{Overall} $\uparrow$  \\
\midrule
  \multirow{6}{*}{MusicXML} & \multirow{3}{*}{\libraryName\--MF} & 10k & 28.6 & 0.98 & 0.90 & 87.0 & -- & -- & -- & -- \\
   &  & 5k & 29.6 & 0.98 & 0.87 & 88.3 & 0.26 $\pm$ 0.34 & 0.30 $\pm$ 0.33 & 0.40 $\pm$ 0.33 & 0.49 $\pm$ 0.34 \\
   &  & -- & 21.2 & 0.98 & 0.85 & 84.5 & \textbf{1.19 $\pm$ 0.38} & \textbf{0.63 $\pm$ 0.34} & 0.41 $\pm$ 0.33 & 0.60 $\pm$ 0.34 \\
   & \multirow{3}{*}{\libraryName\--PF} & 10k & 20.9 & 0.96 & 0.79 & 85.6 & -- & -- & -- & -- \\
   &  & 5k & 26.5 & 0.94 & 0.84 & \textbf{89.2} & 0.09 $\pm$ 0.34 & -0.07 $\pm$ 0.33 & 0.22 $\pm$ 0.33 & 0.19 $\pm$ 0.34 \\
   &  & -- & 19.3 & 0.99 & 0.76 & 78.0 & 0.60 $\pm$ 0.35 & 0.60 $\pm$ 0.34 & \textbf{0.89 $\pm$ 0.36} & \textbf{0.81 $\pm$ 0.36} \\
\midrule
  \multirow{3}{*}{MIDI} & \multirow{3}{*}{REMI+} & 10k & 41.6 & 0.94 & 0.96 & 81.5 & -- & -- & -- & -- \\
   &  & 5k & 46.9 & 0.94 & 0.96 & 79.4 & -0.70 $\pm$ 0.36 & -0.49 $\pm$ 0.34 & -0.23 $\pm$ 0.33 & -0.50 $\pm$ 0.35 \\
   &  & -- & 40.4 & 0.95 & 0.97 & 77.4 & 0.54 $\pm$ 0.35 & 0.52 $\pm$ 0.34 & 0.20 $\pm$ 0.33 & 0.51 $\pm$ 0.34 \\
\midrule
  \multirow{3}{*}{ABC} & \multirow{3}{*}{Word-level} & 10k & 19.5 & 0.99 & 0.98 & 52.6 & -- & -- & -- & -- \\
   &  & 5k & 19.8 & 0.98 & 0.96 & 56.0 & -0.46 $\pm$ 0.34 & -0.31 $\pm$ 0.33 & -0.57 $\pm$ 0.34 & -0.69 $\pm$ 0.35 \\
   &  & -- & 16.6 & 0.98 & 0.97 & 50.6 & -1.46 $\pm$ 0.42 & -1.18 $\pm$ 0.39 & -1.32 $\pm$ 0.40 & -1.40 $\pm$ 0.42 \\
\bottomrule
\end{tabular}%
}
\caption{Generative evaluation across formats and tokenisation strategies. Statistical metrics characterise the generated music distribution; JS measures similarity to the target corpus. User study columns report Bradley--Terry preference scores.}
\label{tab:generative_results}
\vspace{-1em}
\end{table*}

To assess the perceptual quality of generated music, we conduct a pairwise preference study following the evaluation protocol proposed in similar studies~\cite{fradet2023bpe,fradet2023impact,long2024pdmx,qian2026pianoroll}.
Pairwise judgments are aggregated using a Bradley--Terry model to estimate latent quality scores and derive a global ranking of the models, as proposed in similar evaluation frameworks~\cite{chiang2024chatbot, kim2025music}. The scores are mean-centered across models for interpretability.
% Following prior work~\cite{chiang2024chatbot, kim2025music}, pairwise judgments are aggregated using a Bradley--Terry model to estimate latent quality scores and derive a global ranking of the models.

\vspace{-1.2em}
\paragraph*{Stimuli.}
We sample 7 tracks at random from the subsets of the PDMX dataset that were not used for training. 
For each track, we generate 10 continuations per model, conditioned on a prompt consisting of the first 4 bars of the score plus 4 additional tokens. 
We then randomly select 3 continuations per prompt for each model, resulting in 21 sampled continuations per model (168 stimuli in total).
The prompt length is chosen to provide sufficient musical context while mitigating the risk of models, particularly those trained with BPE, prematurely terminating generation by producing an \texttt{EOS} token. 
Participants listen to a synthesised rendering of the prompt followed by 12 generated bars from the two models, presented in randomised order. 
% Note that by design, part-first and ABC models, which serialise parts sequentially, receive as input only the first 4 bars of the first part rather than a multi-part context, which affects the nature of the continuation task for these models.

\vspace{-1.2em}
\paragraph*{Evaluation dimensions.}
For each pair of continuations, participants indicate their preference along four dimensions:
\begin{inlinelist}
    \item \textbf{Fidelity}: consistency with the prompt in terms of pitch, scale, and rhythm;
    \item \textbf{Correctness}: quality of note succession and harmonic content;
    \item \textbf{Diversity}: variety of melodies and harmonies while remaining musically coherent;
    \item \textbf{Overall preference}: subjective global preference.
\end{inlinelist}
Musical competency is assessed via the Gold-MSI questionnaire~\cite{mullensiefen2014goldmsi}.
% , allowing results to be stratified by participants' level of musical expertise.

\vspace{-1.2em}
\paragraph*{Design.}
We adopt a balanced pairwise design where model pairs are sampled to ensure comparable exposure, restricting comparisons to the same prompt. To limit participant fatigue, each model is represented by a single BPE variant per condition: the no-BPE model and the 5k BPE model, yielding 8 models.
Each participant evaluates 30 randomly assigned comparisons. The study involved \participantsNumber\ participants, yielding a total of 960 judgments. On average, each model appears in $\sim120$ comparisons, and each stimulus is evaluated 5--6 times by different participants.
Participants reported an average Gold-MSI score of 5.5 out of 7, indicating a musically experienced sample.
\section{Results}
\label{sec:results}

\subsection{Understanding Tasks}
% Table~\ref{tab:downstream_results} reports the results of the probing evaluation across both understanding tasks. 
% For composer classification, \libraryName\ models consistently outperform both baselines, with MIDI representations ranking second. 
% For difficulty estimation, MusicXML models again achieve the best results; interestingly, ABC representations rank second, suggesting that notational richness plays a meaningful role in this task. 
% Across both tasks, the measure-first serialization consistently yields the strongest results among MusicXML configurations, and the 10k vocabulary performs slightly better than no-BPE and on par with 5k, suggesting a mild benefit from sub-token merging for understanding tasks.
Table~\ref{tab:downstream_results} reports the results of the probing evaluation across all understanding tasks.
For composer classification, \libraryName\ models consistently outperform both baselines, with MIDI representations ranking second.
For difficulty estimation, MusicXML models again achieve the best results; interestingly, ABC representations rank second, suggesting that notational richness plays a meaningful role in this task.
For both tasks, BPE merges provide a consistent benefit, with the 10k vocabulary performing slightly better than no-BPE and on par with 5k, consistent with~\cite{fradet2023bpe}. 
For accompaniment suggestion, \libraryName\ models again rank highest, though BPE does not improve performance here, suggesting that retrieval benefits from fine-grained token resolution rather than compressed representations.
\ap{Notably, the 5k BPE vocabulary outperforms the 10k for MusicXML configurations, suggesting that the larger vocabulary is over-parameterised given the size of the training set.}
% ; unlike the other two tasks no-BPE models consistently outperform their BPE counterparts across all configurations.
% Across all tasks, the measure-first serialisation consistently yields the strongest results among MusicXML configurations.
% , suggesting a mild benefit from sub-token merging for understanding tasks.
Octuple underperforms relative to other tokenisers, contrary to~\cite{zeng2021musicbert,fradet2023bpe}, likely because its pooled embeddings do not account for MusicXML's additional expressive tokens.

The ablation confirms that notation matters: when scores are first converted to MIDI before tokenisation, MIDI-derived \libraryName\ models match the performance of their MIDI counterparts, demonstrating that the advantage of MusicXML lies in the notational content lost upon conversion to MIDI, not in the tokenisation 
strategy alone.

\subsection{Generative Tasks}

Table~\ref{tab:generative_results} reports the results of the generative evaluation across formats, tokenisers, and BPE vocabulary sizes.

\vspace{-1.5em}
\paragraph*{Objective evaluation.}
% Considering the JS score as the primary indicator of distributional similarity to the target corpus, \libraryName\ models consistently outperform both baselines across BPE configurations, with the part-first serialisation at 5k BPE achieving the best overall score (JS = 89.2), followed closely by measure-first at the same vocabulary size (JS = 88.3). 
\ap{\libraryName~models outperform both baselines on JS score across BPE configurations, with part-first at 5k BPE achieving the best result (89.2), followed by measure-first (88.3).}
% The best-performing configuration overall is \libraryName\--PF with a 5k BPE vocabulary (JS = 89.2), followed closely by \libraryName\--MF with the same vocabulary size (JS = 88.3). 
REMI+ occupies an intermediate position, while on average producing more polyphonic output than its MusicXML and ABC counterparts. 
ABC word-level tokenisation performs worst across all configurations, suggesting that word-level tokenisation of ABC notation leads to a distribution that diverges substantially from the reference. 
Notably, the 5k BPE vocabulary consistently outperforms the 10k, likely due to vocabulary over-parameterisation given the size of the training set. 
% A secondary observation concerns the polyphony rate: MIDI models produce markedly more polyphonic output than their MusicXML and ABC counterparts, reflecting an inherent representational difference between performance-oriented and score-based formats rather than a quality differential per se.

% Overall, these results suggest that MusicXML-based tokenization yields generated sequences whose harmonic, rhythmic, and tonal characteristics more closely match the target distribution than either MIDI or ABC representations.

\vspace{-1.5em}
\paragraph*{User study.}  The Bradley--Terry scores corroborate the statistical findings, with ABC tokenisation ranking lowest and MIDI models scoring slightly below their MusicXML counterparts. 
Among MusicXML configurations, part-first achieves the highest overall preference (0.81) and diversity (0.89) scores, while measure-first leads on fidelity (1.19) and correctness (0.63), suggesting that the two serialisation orders capture complementary generative strengths. 
% Among MusicXML configurations, \libraryName\--PF without BPE achieves the highest overall preference (0.81) and diversity (0.89).
% \libraryName\--MF without BPE leads on fidelity (1.19) and correctness (0.63).
Contrary to~\cite{fradet2023bpe}, BPE does not improve perceptual quality here, which we attribute to a methodological difference: whereas~\cite{fradet2023bpe} apply BPE at the bar level, we apply it over the full sequence, which may break merge boundaries at the prompt and yield suboptimal continuations.
% Contrary to findings in prior work~\cite{fradet2023bpe}, BPE does not improve perceptual quality here. 
% We attribute this to a methodological difference: whereas~\cite{fradet2023bpe} apply BPE at the bar level, we apply it over the full sequence for consistency across tokenisers, which may break BPE merges at the prompt boundary and yield suboptimal continuations.
% ; however, this result should be interpreted with caution, as part-first models receive only the first four bars of a single part as prompt, rather than a multi-part context. 
% This input truncation artificially inflates diversity while making the continuation task easier, and therefore does not straightforwardly reflect model quality. 
% , suggesting that measure-first serialization produces output perceived as more faithful and notationally coherent.
\section{Conclusions}
\label{sec:conclusions}

We presented \libraryName, the first general-purpose tokenisation framework for MusicXML, addressing a representational gap that has long limited the applicability of deep learning models to score-based symbolic music. 
% To assess whether preserving score semantics at the tokenization level benefits downstream modelling, we evaluated \libraryName\ against MIDI and ABC baselines on both understanding and generative tasks.
Our empirical study demonstrates that this gap has measurable consequences: MusicXML-based tokenisation matches or outperforms the baselines.
% Across all settings, MusicXML-based tokenization consistently matches or outperforms the baselines. 
% On understanding tasks, \libraryName\ models achieve the best results on composer classification and difficulty estimation, suggesting that notational expressiveness is beneficial in such tasks. 
% On the generative side, \libraryName\ models produce output that more closely matches the target distribution and is preferred by listeners in perceptual evaluation. 
% Together, these results indicate that score semantics carry information that performance-oriented and plain-text representations fail to capture, and that this gap has measurable consequences for both what models learn and what they generate.
On understanding tasks, \libraryName\ models achieve the best results on composer classification, difficulty estimation, and accompaniment suggestion, indicating that notational expressiveness is beneficial in such tasks.
On the generative side, \libraryName\ models produce output that more closely matches the target distribution and is preferred by listeners in perceptual evaluation.
Among serialisation strategies, measure-first consistently outperforms part-first on understanding tasks, while part-first is preferred in generative evaluation, suggesting that the two orders capture complementary inductive biases.
Together, these results indicate that score semantics carry information that performance-oriented and plain-text representations fail to capture, and that this gap has measurable consequences for what models learn and generate.

% \newpage

\section{Acknowledgments}

\ap{This work is supported by the ``Cátedra IA y Música'' project (TSI-100929-2023-1), funded by the Secretaría de Estado de Digitalización e Inteligencia Artificial, the European Union-Next Generation EU funds and BMAT Music Innovators. And by the ``IMPA'' project (PID2023-152250OB-I00) funded by MCIU/AEI/10.13039/501100011033/FEDER, UE.}

% \section{AI Usage Statement}

% In line with the ISMIR 2026 AI Usage Policy, if you need to declare and explain the use of AI tools in the work described in the paper, you can do so by including a section titled ``AI Usage Statement" at the end of the paper. The AI Usage Statement section can be included both at submission time and in your camera-ready version. See the Call for Papers and Author Guidelines for more details. The AI Usage Statement section does \textit{not} count towards the page limit for scientific content. 

\section{Ethics Statement}

This work involves training generative models on symbolic music data, an area that raises legitimate concerns regarding the rights of composers and musicians, and the broader societal implications of automated music generation.
% We are mindful of these concerns and have taken deliberate steps to ensure that our experimental setup is ethically sound.
All models in this study are trained exclusively on PDMX (\textit{v5})~\cite{long2024pdmx}, a dataset comprising scores released under CC-0 or Public Domain licenses.
% , with no copyrighted material involved at any stage of training or evaluation.
\ap{We note, however, that such licences may not always reflect the status of the underlying composition; a portion of the corpus is therefore likely to consist of arrangements of works still under copyright. We rely on the licence metadata as the best available signal, and flag this as a limitation of currently available large-scale symbolic corpora.}

% For BibTeX users:
\bibliography{references}

\end{document}